# Stoichiometry and Thickness of Epitaxial $SrTiO_3$ on Silicon (001): an Investigation of Physical, Optical and Electrical Properties

Andries Boelen[1,2], Marina Baryshnikova[1], Anja Ulrich[1,3], Kamal Brahim[1,4], Joris Van de Vondel[5], Christian Haffner[1] and Clement Merckling[1,2]

1) Imec, B-3001 Leuven, Belgium
2) Department of Materials Engineering (MTM), KU Leuven, B-3001 Leuven, Belgium
3) Department of Information Technology (INTEC), Photonics Research Group, Ghent University, B-9052 Ghent, Belgium
4) Department of Electrical Engineering (ESAT), KU Leuven, B-3001 Leuven, Belgium
5) Department of Physics and Astronomy, KU Leuven, B-3001 Leuven, Belgium

## Abstract

Strontium titanate ($SrTiO_3$, STO) stands out as a promising material for various electronic applications thanks to its exceptional dielectric properties. Molecular beam epitaxy is one of the few techniques which allows epitaxial growth of STO directly on industry-relevant silicon substrates. However, maintaining precise stoichiometry and high crystalline quality in this process remains a significant challenge. Establishing this is essential to obtain STO with bulk-like dielectric properties and to minimize leakage current and optical absorbance. In this study, the importance of cationic stoichiometry and the effect of thickness are investigated for STO thin films epitaxially grown on silicon. We employed real-time reflection high-energy electron diffraction (RHEED) as a feedback loop mechanism to counteract Sr source oxidation and maintain a constant flux. Additionally, high-temperature post-growth annealing treatments in $O_2$ were investigated to promote layer relaxation and reduce oxygen vacancy concentration, thereby improving the physical, electrical, and optical properties of stoichiometric STO. As a result, high-quality STO thin films exceeding 100 nm were successfully fabricated featuring a bulk-like out-of-plane lattice parameter and refractive index, as well as rocking curve full width at half maximum below 0.2°, smooth surface ($R_q < 0.2$ nm) and a leakage current density below 1E-7 A/cm$^2$.

## I Introduction

High permittivity materials garnered significant interest for their potential application as gate dielectrics in CMOS technology, prompting extensive research into their exploration and performance enhancement [1]. At room temperature strontium titanate ($SrTiO_3$, STO) possesses a high dielectric permittivity, good insulating properties and although paraelectric, it can be altered into a ferroelectric state by i.a. doping [2,3] or strain engineering [4]. Therefore, it could be used in various essential technologies such as high-density non-volatile memory [5], nanocapacitors used as batteries [6] and electric field sensors for medical imaging and security [7].

The relative dielectric permittivity of bulk STO reaches values of 20,000 and above at cryogenic temperatures thanks to its quantum paraelectric state [8]. This behavior is caused by quantum fluctuations of the $Ti^{4+}$ atoms, which prevent the ferroelectric phase transition. STO's quantum paraelectric behavior, and thus large permittivity value, is sustained even at high radio frequencies (RF) due to its phononic lattice resonances. Both efficient electrical energy storage (large permittivity) and fast data transmission (high bandwidth) make STO an interesting candidate for information technologies operated at cryogenic temperatures, such as quantum computing. For instance, according to first principle calculations, the high permittivity should result in a large Pockels coefficient r for strained STO, which could be used for light switching technologies [9]. Recently, $r_{eff} > 400$ pm / V has been reported for bulk STO due to the quadratic electro-optic Kerr effect [10]. However, literature on STO thin film permittivity is limited and results report maximal values of only ~ 1,000 when approaching 0 K [11,12]. This shortcoming can be attributed to a higher density of defects and dislocations, arising from e.g. stoichiometry instabilities, altering the crystal structure. Secondly, strain, due to a lattice mismatch with the substrate, will induce a ferroelectric phase transition at cryogenic temperatures [13]. Larger strain elevates the Curie temperature $T_C$, (maximal permittivity), thereby reducing the permittivity at 0 K. Consequently, minimal strain is preferred for cryogenic applications. In this work, we aim to address these issues related to material quality.

Epitaxial STO thin films with perfect stoichiometry can be grown using RF sputtering [14] and pulsed laser deposition (PLD) [15] techniques. However, the best quality is achieved on oxide substrates. For scalability and integration, high-quality monocrystalline STO films have to be deposited on silicon substrates. This limits suitable techniques to molecular beam epitaxy (MBE), which can ensure a stable, well-defined Si/STO interface [16] essential for epitaxial growth.

MBE allows uniform deposition of high-quality films on silicon with minimal grain boundaries, crucial for applications needing exceptional electronic and optical properties [17,18]. Furthermore, MBE allows atomically precise control over film thickness and stoichiometry with independent fluxes. It also supports smooth surfaces and abrupt interfaces, necessary for advanced device fabrication. One of the biggest challenges of MBE of thick STO layers though is the source oxidation during growth [19]. This means that if no special care is taken to neutralize the flux depletion, the final cationic stoichiometry of the STO layer will deviate from the desired Sr/Ti = 1. On top of that, as-grown layers often suffer from oxygen deficiency due to UHV conditions of MBE and require additional annealing in $O_2$ atmosphere.

Most research focuses on 5 – 15 nm thin films of STO, used not for its unique properties but rather as a buffer layer for growing other perovskites such as $BaTiO_3$ [20,21] or $LaAlO_3$ [22]. Also in our previous study [23] we looked in depth in the relaxation behavior of 15 nm STO templates on Si with varying cationic stoichiometry, aiming to improve their quality. Because such thin layers are strongly subjected to strain and interface effects, their overall properties will not match those of the bulk crystal.

Therefore, the purpose of this study was to establish a process for MBE growth of high-quality thick (> 50 nm) STO layers on silicon (001)-oriented substrates supported by physical, electrical and optical characterization. Since perfect stoichiometry is crucial for obtaining STO films with bulk-like properties, the impact of both Sr and Ti excess was investigated. Source oxidation during oxide epitaxy is the main cause of uncontrollable stoichiometry and a method to obtain Sr/Ti = 1 was established. Moreover, post-growth annealing (PGA) treatments in $O_2$ atmosphere were performed to reduce the oxygen vacancy concentration inside the layers, also inducing strain relaxation. Lastly, the STO thickness was varied to compare strain effects as well as the influence of interfacial $SiO_2$ for varying thicknesses of stoichiometric STO.

## II Experimental

Heteroepitaxial STO thin films were grown on silicon (001)-oriented substrates using a 200 mm Riber 49 MBE tool, which operates at a base vacuum of 1E-10 Torr. This system was equipped with reflection high-energy electron diffraction (RHEED) to monitor the surface crystal structure in real-time. Strontium was evaporated from a dual-filament Knudsen effusion cell while the titanium molecular beam was generated by an electron beam evaporator. The flux of both metallic molecular beams was calibrated in-situ using a quartz crystal microbalance (QCM). For stoichiometric STO, the Sr and Ti fluxes were 1.3 Å/s and 0.6 Å/s, respectively. Molecular and/or atomic oxygen were introduced to the MBE chamber through a radio frequency (RF) plasma source. The $O_2$ flow was maintained at 5 sccm by a mass flow controller (MFC).

The p-type Si (001) wafers were cleaned for 90 s in a 2% HF solution to remove part of the organic residues from the surface prior to ultra-high vacuum (UHV) introduction. Once in the MBE growth chamber, they were first heated, and a thin Sr layer (~ 1 nm) was deposited (680°C) for further Sr-assisted native oxide desorption (730°C) [24]. This led to a slight (3 x 2) surface reconstruction when the substrate was cooled down to 500°C [25]. At this temperature the Sr interfacial layer was completed, achieving ½ monolayer that forms an oxidation barrier between Si and STO. This was confirmed by a (2 x 1) surface reconstruction on the RHEED pattern. After native oxide removal, direct epitaxy of the first 3 nm of STO, with [100] STO (001) // [110] Si (001) [16], was performed in molecular oxygen at 350°C. After this, growth was paused to switch to atomic oxygen (500 W plasma power) into the growth chamber and to increase the substrate temperature to 550°C. Under these conditions, the remaining STO epitaxy was completed at a growth rate of approximately 1 nm/min (~ 9E-7 Torr). Moreover, since from our previous study [23] we know that as-grown MBE STO layers suffer from oxygen deficiency, a cooling down step to 200°C was performed over 45 min in oxygen atmosphere (~ 2E-6 Torr), acting as an in-situ annealing treatment, to improve oxygen stoichiometry. An impact of this modification is discussed further in this paper.

Cation stoichiometry (Sr/Ti ratio) was determined by Rutherford backscattering spectrometry (RBS) using a 1.5 MeV He+ ion beam. The crystalline state of the films was examined by high-resolution X-ray diffraction (HR-XRD) using a PANalytical X'pert Pro diffractometer with Cu $K_{\alpha 1}$ radiation line ($\lambda \approx 1.54$ Å). The structural properties were measured by symmetric 2θ-ω and ω scans around the STO (002) Bragg diffraction peak and aligned with respect to the Si (004) peak. The STO surface roughness and morphology were investigated using a Bruker Dimension Icon atomic force microscopy (AFM) tool in pulsed force mode.

Transmission electron microscopy (TEM) was used along with focused ion beam (FIB) for sample preparation to observe the microstructure and chemical composition. A protective spin-on-carbon (SOC) layer was used to cap the lamella. The Titan3 G2 tool was operated at 200 kV to perform high-angle annular dark-field scanning transmission electron microscopy (HAADF-STEM), annular bright-field STEM (ABF-STEM), dark-field STEM (DF-STEM), and energy-dispersive X-ray spectroscopy (EDS).

Ex-situ post-growth annealing (PGA) treatments in $O_2$ at atmospheric pressure were carried out using an Annealsys rapid thermal annealing system AS-One. Spectroscopic ellipsometry (SE) was performed using a J.A. Woollam RC2 tool. Data were collected at angles between 50° and 80° with steps of 5° using an acquisition time of 3 s. Accessory CompleteEASE software was used to fit and extract the spectra of refractive index $n$ and extinction coefficient $k$, with the model described in Appendix B.

To form metal contacts, the front and back sides of the samples were sputtered with 90 nm of Cr and Pt, respectively. Using a shadow mask, the top Cr contacts form circular dots with a radius of 200 µm. A Keithley 4200-SCS parameter analyzer connected to a Summit 11000 probing station measured the out-of-plane leakage current through the stack. The Pt back contact was grounded, while a Cr contact dot was biased from 0 V to −4 V and back to 0 V, and then from 0 V to +4 V and back, in steps of 10 mV with a delay of 0.3 s.

## III Results

*Source oxidation*

When introducing oxygen into the MBE growth chamber during the STO epitaxy both Sr and Ti sources will be gradually oxidized, resulting in the formation of SrO and $TiO_2$ at their surface causing a flux reduction over time (see Figure 1) [19]. An "automated" feedback loop with a cross beam quadrupole mass spectrometer (XBS) keeps the Ti atomic flux stable. However, this system is not compatible with effusion sources like Sr, leading to an unadjusted Sr molecular beam and consequently resulting in Ti-rich STO films (Sr/Ti < 1).

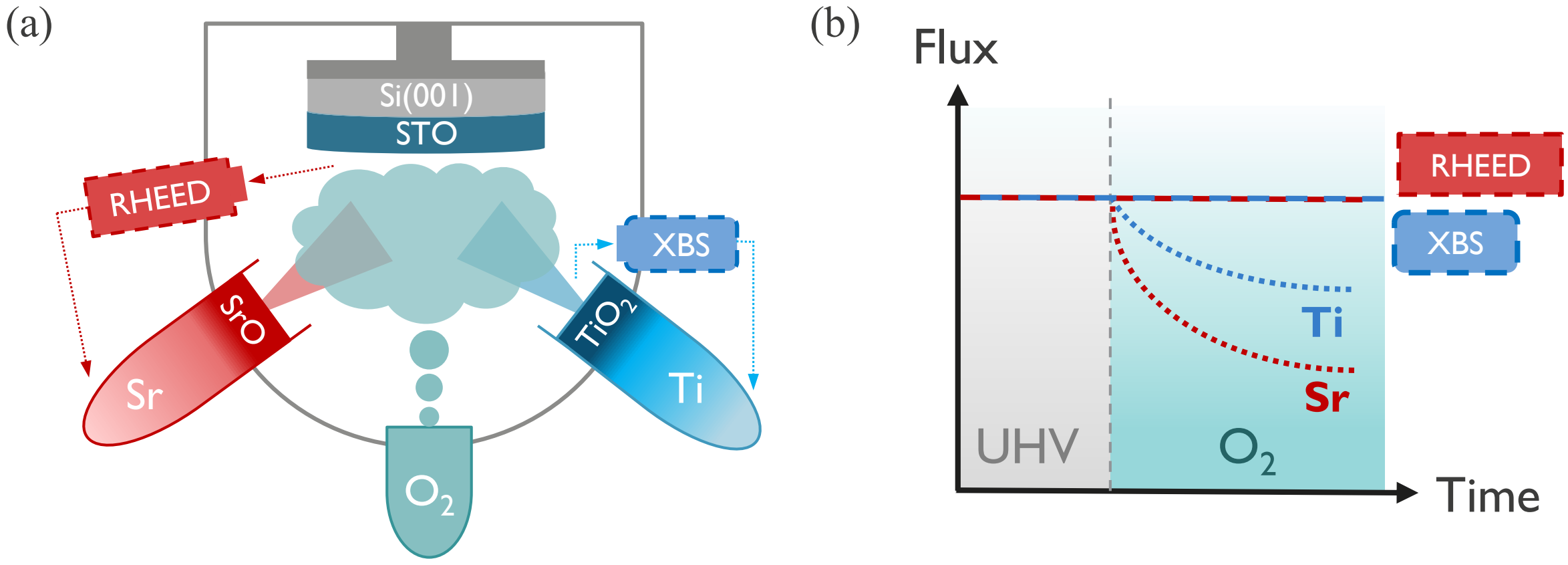

*Figure 1: (a) Schematic of the MBE growth chamber consisting of the atomic sources with their feedback loop mechanisms. (b) Using RHEED and XBS to maintain stable Sr and Ti beam fluxes, respectively, to neutralize the reduction of metallic beam fluxes due to source oxidation.*

The Sr source oxidation effect can be counteracted by manually increasing the Sr source temperature during evaporation. For this we use in-situ RHEED to monitor in real-time the crystalline state of the STO surface. The aim is to have a sharp high-contrast diffraction pattern without any surface reconstruction, such as in Figure 2b, which indicates the growth of high-quality stoichiometric (1x1) STO (001). A x2 surface reconstruction along the $[100]_{STO}$ direction reveals a (2x1) $Ti_2O_3$ crystal top layer as in Figure 2a, which will cause the STO layer to end up being Ti-rich if the reconstruction proceeds through the full film. Similarly, a x2 surface reconstruction along the $[110]_{STO}$ direction as in Figure 2c indicates a c(2x2) SrO structure, eventually resulting in a Sr-rich STO film [26]. The surface reconstructions can be seen more clearly on the corresponding RHEED intensity line profiles. In this regard, RHEED is used here as a feedback loop for maintaining a constant Sr flux and obtaining stoichiometric films. Without the use of RHEED, our STO 50 nm films would typically end up non-stochiometric with 0.70 < Sr/Ti < 0.80.

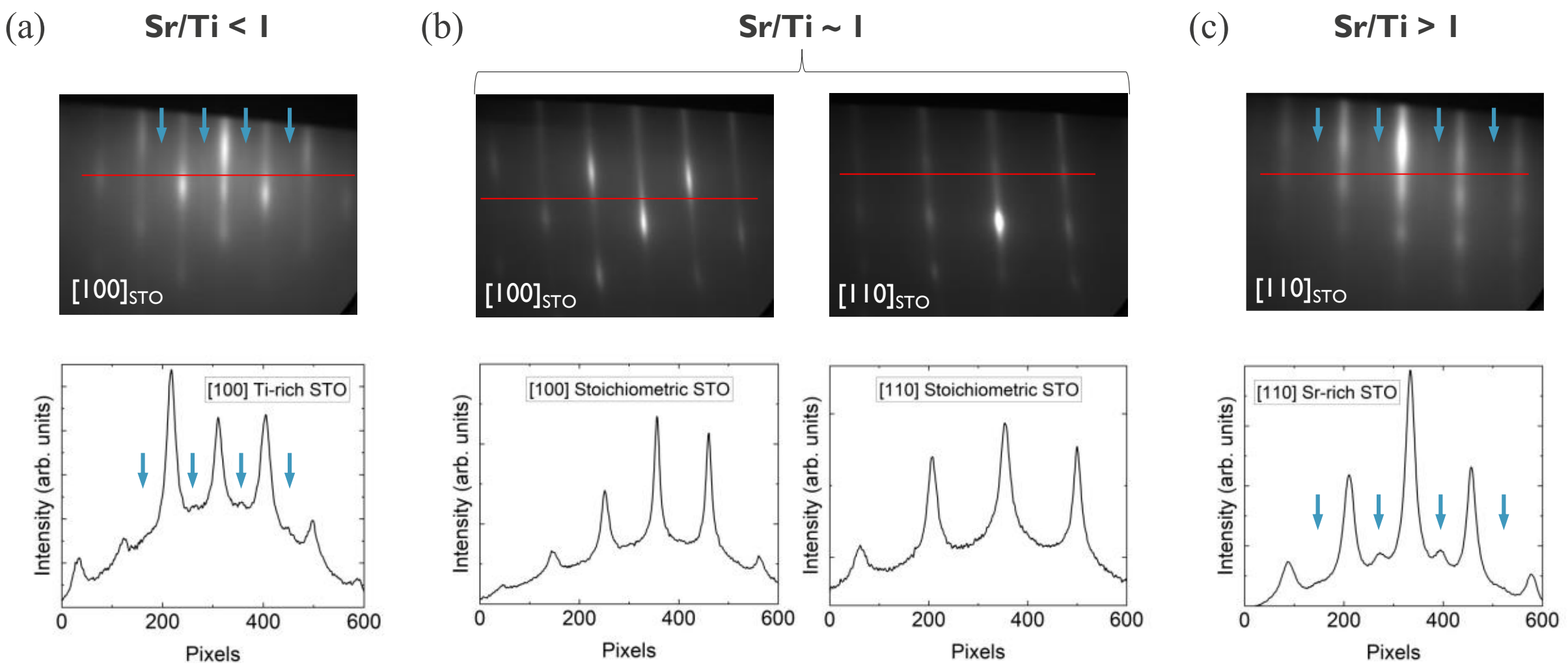

*Figure 2: Real-time RHEED patterns with corresponding line profiles during STO epitaxy. (a) A x2 surface reconstruction (see blue arrows) along $[100]_{STO}$ indicates Ti-rich STO. (b) A sharp high-contrast RHEED pattern along both $[100]_{STO}$ and $[110]_{STO}$ indicates stoichiometric STO. (c) Oppositely, a x2 surface reconstruction along $[110]_{STO}$ indicates a Sr-rich STO layer.*

*Stoichiometry impact*

STO thin films have been grown on Si with various thicknesses, growth temperatures ($T_G$), and cationic (Sr/Ti) stoichiometries. Figure 3 summarizes the out-of-plane lattice parameter (c) measured by XRD as a function of the Sr/Ti ratio determined by RBS. The majority of the STO samples on the graph are Ti-rich, this is due to the oxidation of the Sr source discussed above. From the presented results it is evident that to match the lattice parameter of bulk STO (c = 3.905 Å) a few conditions need to be met: a Sr/Ti ratio close to 1 together with a high growth temperature and layer thickness above ~ 50 nm. This can be explained by the fact that higher growth temperatures and layer thicknesses promote relaxation of compressively strained films of STO on Si [27,28]. At the same time, a perfect stoichiometry is required to minimize the lattice parameters and overall cell volume, as it was already reported earlier by other researchers [29].

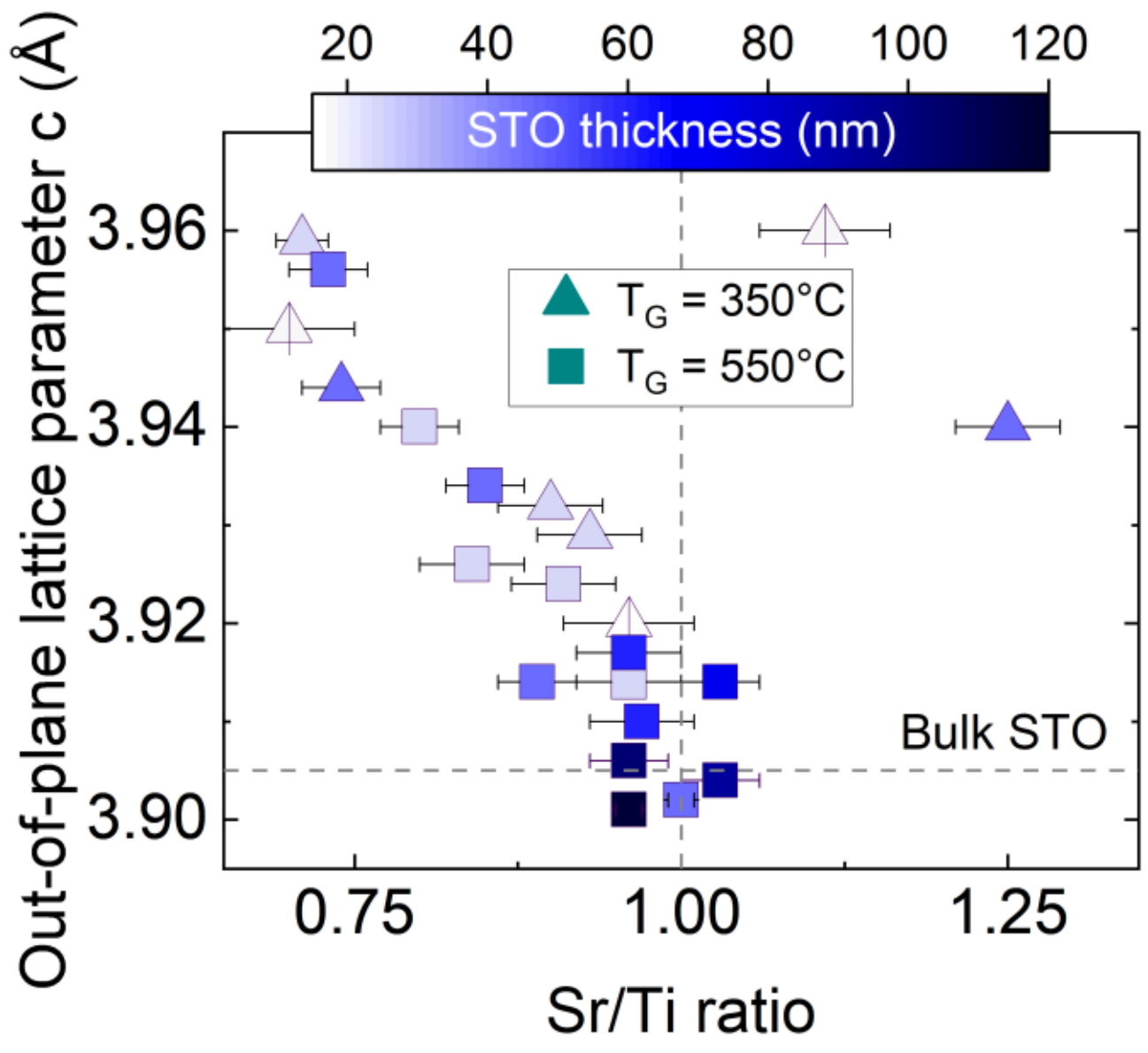

*Figure 3: Out-of-plane lattice parameter as a function of Sr/Ti ratio for STO films with varying thickness and growth temperature $T_G$. The three thinnest films (triangles with vertical line) are taken from our previous work* [23].

To further investigate the impact of Sr/Ti non-stoichiometry, caused by flux variations during growth, three different samples with a target thickness of 50 nm were grown at 550°C with a different cationic stoichiometry: Sr/Ti = 0.78, 1.02 and 1.46. They are referred to as Ti-rich, stoichiometric and Sr-rich STO samples, respectively. These three STO samples were subjected to high-temperature PGA treatments in $O_2$ at 850°C for 30 min and 60 min to find the best conditions for the formation of fully relaxed and oxidized STO.

Figure 4a and 4b show 1 μm x 1 μm AFM images of the three samples with different Sr/Ti ratio before and after PGA at 850°C for 60 min. The surface of all as-grown samples looks smooth ($R_q$ < 0.4 nm) and it's only on the stoichiometric sample where small islands of ~ 6 nm height are visible. Applying PGA improved the surface of the stoichiometric sample ($R_q$ = 0.18 nm) by reducing the density of the outgrowths and decreasing their height to ~ 2 nm. At the same time a formation of steps with a height of 1 unit cell (~ 0.4 nm) is observed, indicating a SrO- or $TiO_2$- terminated STO surface. High-temperature annealing had an overall negative effect on samples with deviations from the Sr/Ti stoichiometry, which is in agreement with previous work [23, 30]. For the Ti-rich STO it promoted ~ 100 nm island formation with a height of ~ 20 nm, while for the Sr-rich sample deep cracks and pinholes occurred. To further understand the reason for such a severe degradation of the Sr-rich sample, a cross-sectional scanning electron microscopy (XSEM) investigation was performed (Figure 4c). This shows the formation of ~ 20 nm voids in STO and a ~ 40 nm interfacial layer between STO and Si, which was not present before annealing. Figure 4d summarizes the root mean square (RMS) surface roughness $R_q$, obtained from 10 x 10 μm AFM (*Supplementary Figure 1*), for the three samples before and after PGA treatment.

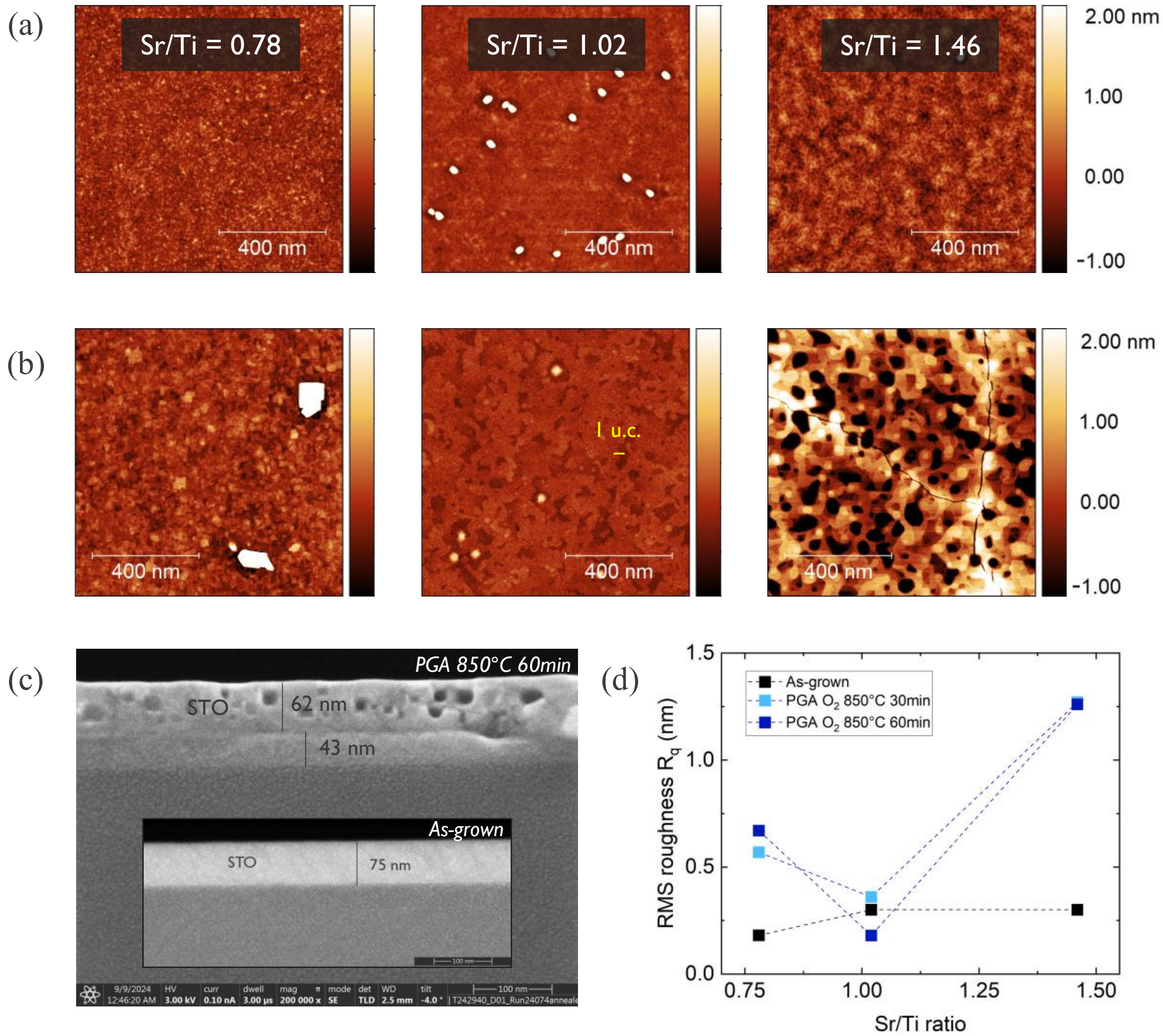

*Figure 4: (a) AFM images of the three as-grown samples with varying stoichiometry. (b) AFM images of the three samples with varying stoichiometry after PGA 850°C 60 min. (c) XSEM of Sr-rich STO after PGA 850°C 60 min. Voids inside the STO layer and a thick interfacial layer cause the cracks and pinholes (AFM) of the degraded STO. The inset shows the uniform layer before PGA. (d) AFM surface roughness as a function of Sr/Ti ratio before and after PGA. After PGA the roughness of non-stoichiometric STO increases.*

The importance of cationic stoichiometry control is further emphasized for the crystalline structure of STO. Figure 5 summarizes such structural properties of the STO films, including the FWHM of the STO (002) rocking curve, the out-of-plane lattice parameter (c) and the intensity of the STO (002) reflection, both before and after PGA. As expected, as-grown stoichiometric STO has the lowest FWHM and out-of-plane lattice parameter, which even matches the STO bulk value, combined with the highest peak intensity. Excess of Ti atoms leads to formation of a slightly larger cell but had very little impact on the FWHM and (002) peak intensity. At the same time, incorporation of extra Sr atoms leads to considerable cell expansion in the out-of-plane direction, an increase of the FWHM value and significant reduction of the (002) reflection intensity. These observations suggest that excessive amount of Ti had a relatively small impact on crystallinity of as-grown STO, whereas addition of extra Sr caused significant deterioration of the layer crystallinity. Additional 2θ-ω scans on these off-stoichiometric samples in the wider range did not reveal presence of any extra peaks that can be attributed to pure $TiO_2$ or SrO phases, this indicates that excess of Ti and Sr is probably incorporated as interstitials or amorphous inclusions [31].

Ex-situ annealing of STO layers in $O_2$ is expected to decrease both the FWHM and out-of-plane lattice parameter due to a reduced oxygen vacancy concentration [32], strain relaxation [27] and improved crystallinity [33]. However, in the current study it only had marginal positive effects on the structural properties of stoichiometric and Ti-rich layers. This indicates that applying an extended cooling down step under oxygen (in-situ annealing) was already sufficient to form fully oxygenated layers [34]. For Sr-rich STO films, however, PGA significantly increased the (002) peak intensity and reduced the FWHM and out-of-plane lattice parameter. These results indicate that the crystalline quality, initially compromised by excessive amount of Sr, can be restored by applying PGA. Nevertheless, looking at the results of XSEM investigation, the effect of PGA can hardly be called positive as it led to severe degradation of the layer morphology. The improvement of crystallinity in this case can be explained by Sr segregation effects at elevated temperatures which is known for perovskites materials [35] and which we also saw for our 15 nm thin templates [23].

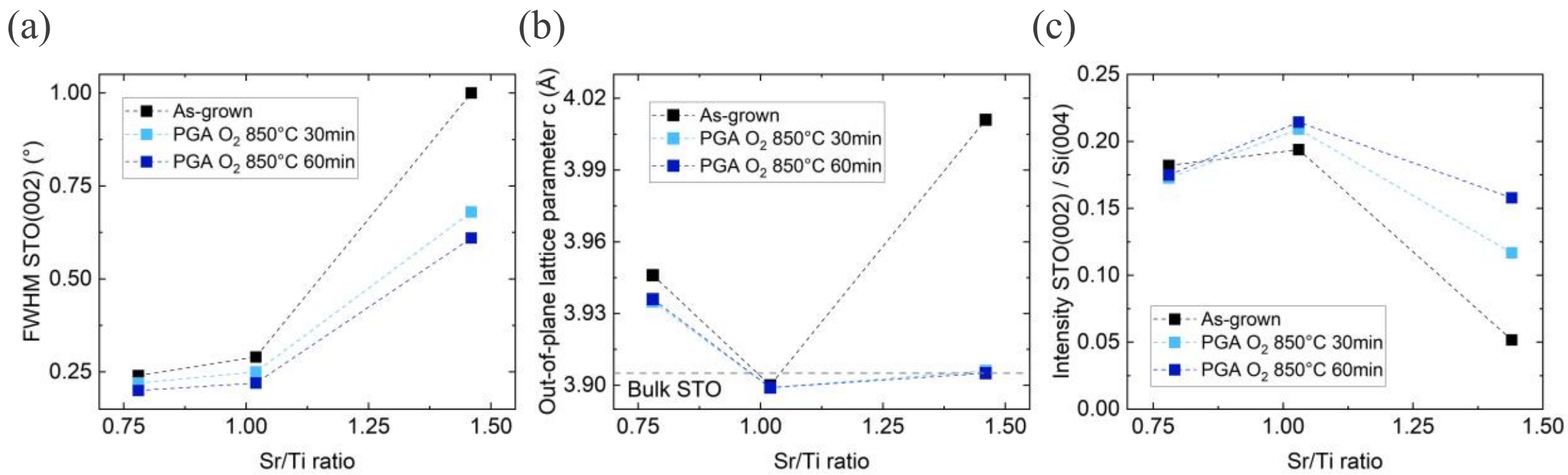

*Figure 5: XRD results for the three STO films with varying stoichiometry before and after PGA. (a) FWHM of the STO (002) rocking curve, which is low for Ti-rich and stoichiometric STO. (b) Out-of-plane lattice parameter, stoichiometric films have a similar lattice parameter as bulk STO. (c) Peak intensity of the STO (002) diffraction peak, maximal for stoichiometric STO after PGA.*

Maintaining precise stoichiometry is essential for the insulating and optical properties of both as-grown and annealed STO films. Figure 6a and 6b display the refractive index (n) and extinction coefficient (k) spectra obtained from ellipsometry measurements. Without PGA, only stoichiometric STO films have a refractive index similar to the bulk value and losses below the sensitivity range of the method [23,36]. PGA has little effect on these values for stoichiometric samples. In contrast, as-grown Ti-rich STO films show different optical characteristics, with the refractive index varying from the bulk value depending on the wavelength and increased losses at wavelengths greater than 300 nm. PGA helps to restore the n and k values of Ti-rich STO films to near-bulk levels. As-grown Sr-rich STO films have a generally lower refractive index probably due to excess of SrO, which has a refractive index around 1.9 [23,37], and increased absorption at wavelengths greater than 300 nm. However, it is challenging to determine if PGA improves the optical properties of Sr-rich STO films due to significant morphology degradation, complicating the optical model for ellipsometry calculations.

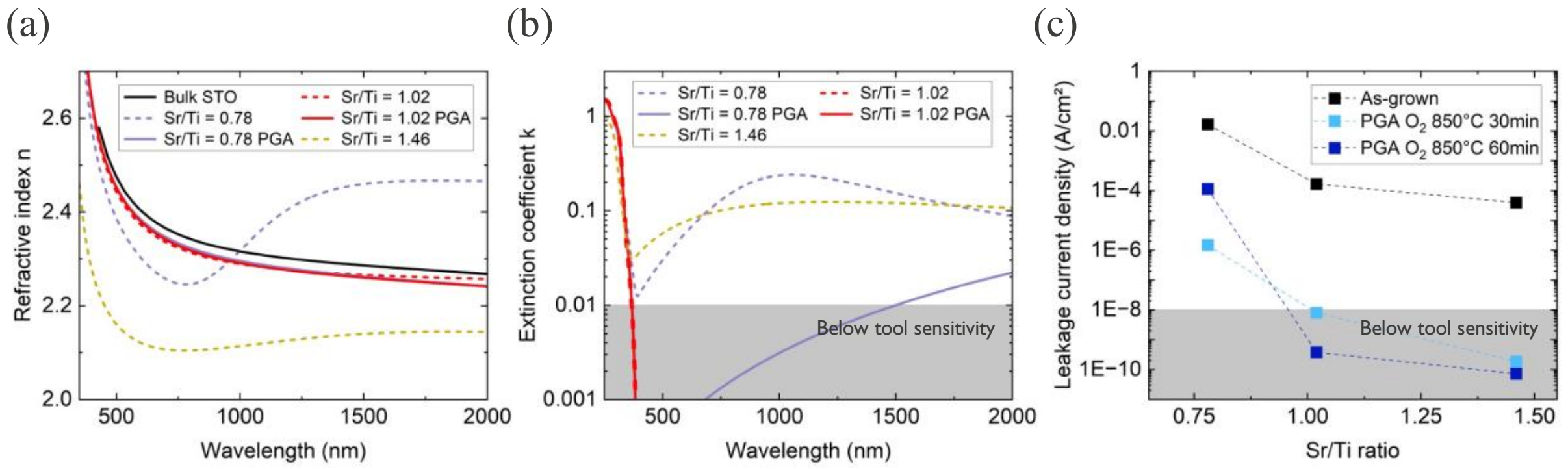

*Figure 6: (a) Refractive index n for the STO layers before and after PGA with bulk reference [38]. (b) Extinction coefficient k for the STO layers before and after PGA showing high optical losses for as-grown non-stoichiometric STO. (c) Out-of-plane leakage current density at 1E+7 V/m as a function of Sr/Ti ratio before and after PGA. Highly Sr-rich STO shows a surprisingly low leakage current, but this cannot be related to STO itself due to the poor crystal quality.*

Figure 6c illustrates the out-of-plane leakage current density before and after PGA for three samples with varying Sr/Ti ratios. The leakage current is minimized when the Sr/Ti ratio is greater than or equal to 1 [23,39,40]. Generally, annealing in oxygen reduces the leakage current by filling oxygen vacancies, which would otherwise introduce free electrons into the conduction band [41]. However, in our measurement setup, the reduction in leakage after annealing can also be attributed to the increase in the $SiO_x$ interfacial layer.

As known from our previous studies [23,42,43], the formation of a thin $SiO_x$ layer at the Si/STO interface is common, resulting from oxygen diffusion into Si at elevated temperatures. This additional layer must be considered when building a reliable optical model for ellipsometry and when discussing leakage current measurements. For our 15 nm STO grown at low temperatures, the interfacial layer thickness ranged from 1 nm to 1.5 nm. In the current study, however, the samples were grown for a longer time and at higher temperatures, therefore a dedicated TEM study was performed to assess how these new conditions affect the layer thickness. TEM results of a different 66 nm STO film in Figure 7 before (left) and after PGA (right) show the STO layer is grown epitaxially even though threading dislocations are present going from interface to the top surface. They also indicate that the $SiO_2$ layer thickness increased to 2.3 nm for the as-grown sample due to different growth conditions and applying PGA further increased it to 4.2 nm. These values have been incorporated into the ellipsometry model together with AFM surface roughness. The presence of this layer will also impact out-of-plane leakage measurements, though additional studies are needed to fully understand its effects.

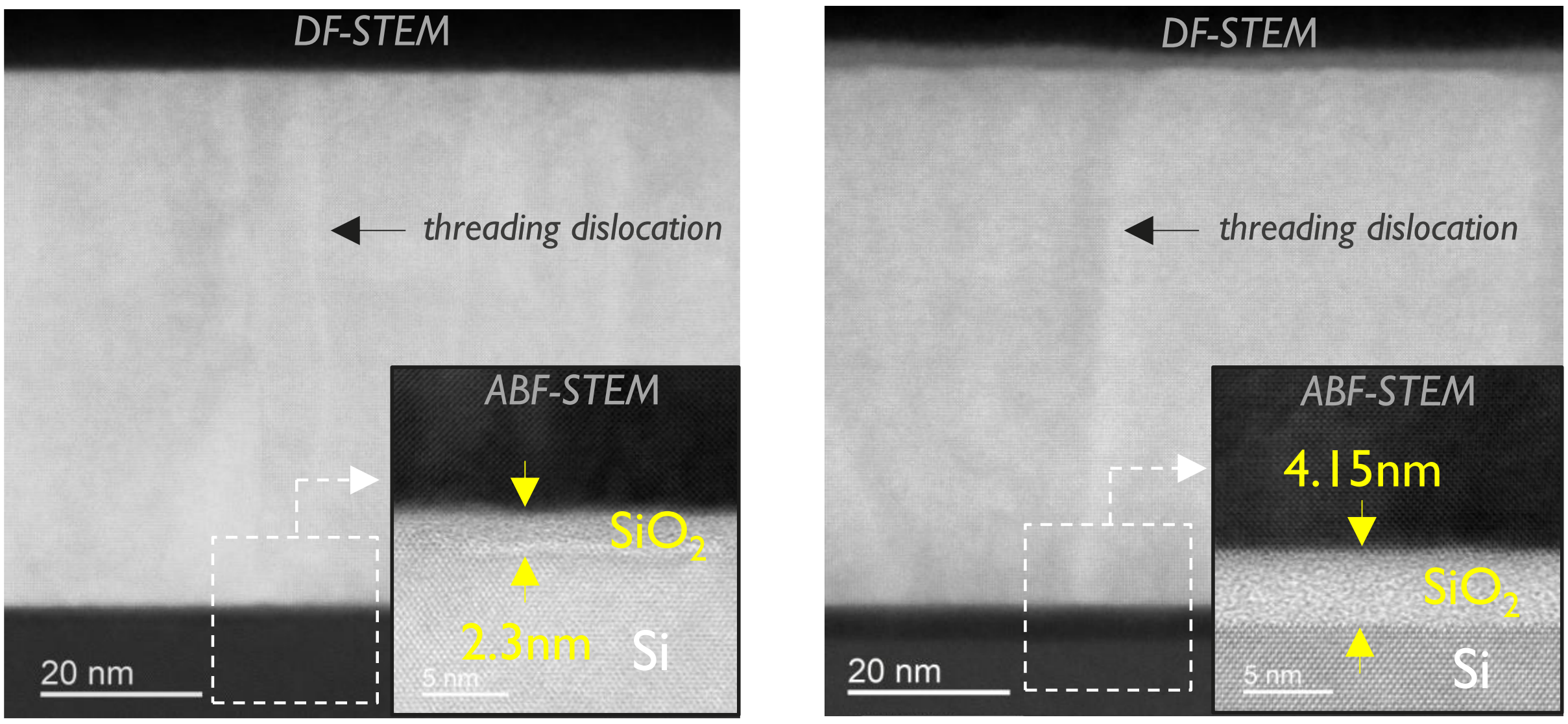

*Figure 7: DF-STEM images of 66 nm STO (Sr/Ti = 1.03) before (left) and after PGA at 800°C for 15 min in $O_2$ (right). The inset shows a zoomed in ABF-STEM image of the interfacial $SiO_2$ layer between STO and Si.*

*Thickness effects*

Now that cationic stoichiometry control has been enhanced, three stoichiometric STO samples with varying thickness have been investigated in a similar way as above. The obtained thicknesses from ellipsometry of 26 nm (Sr/Ti = 0.96), 50 nm (Sr/Ti = 1.02) and 103 nm (Sr/Ti = 0.96) confirm the STO thickness control with corresponding target values of 25 nm, 50 nm and 100 nm.

Figure 8a and 8b illustrate the changes in the FWHM of the STO (002) rocking curve and the out-of-plane lattice parameter with varying thickness and high-temperature PGA in oxygen. The FWHM for as-grown STO films decreases by approximately threefold as the thickness increases from 25 nm to 100 nm. Concurrently, the out-of-plane lattice parameter also decreases. These observations suggest a relaxation process occurring in the layers as thickness increases [27,28]. PGA treatment further reduces the FWHM, particularly in thinner samples, but does not significantly affect the out-of-plane lattice parameter. The lack of reduction in the out-of-plane lattice constant after annealing, especially in thin samples, was unexpected. Since lattice expansion in STO can result from strain or oxygen deficiency, these findings suggest that cooling in oxygen plasma may already introduce sufficient oxygen to form stoichiometric layers, at least in thicker STO. However, this does not explain why the out-of-plane lattice parameter for thin STO does not decrease, necessitating further investigation.

The results of leakage current measurements and optical investigations are presented in Figure 8c and 8d. The leakage current density shows a gradual increase with the thickness of the as-grown STO layer. Ex-situ PGA effectively reduces the leakage current, particularly in the 25 nm and 50 nm thick samples. Since leakage mechanisms in perovskite oxides are often associated with oxygen vacancies [23,33,41], this trend may be attributed to a higher oxygen deficiency in thicker STO specimens. However, the presence of interfacial $SiO_x$, whose thickness is influenced by the growth and PGA conditions, prevents definitive conclusions. As such, these findings should be interpreted with caution, and further studies are necessary to clarify the effects of thickness and annealing on the leakage current.

Despite the inconclusive results regarding leakage current, ellipsometry measurements indicate that the growth approach used, combined with ex-situ PGA, allows for the growth of layers with excellent optical properties regardless of thickness. The spectra of the refractive index *n* for different thicknesses after PGA are shown in Figure 8d. They show behavior very similar to the bulk reference for all three thicknesses, indicating overall high crystalline quality [44]. The minor thickness dependence can be attributed to the proximity of the interfacial $SiO_2$ ($n_{633nm}$ ~ 1.4) for thinner films [42,45].

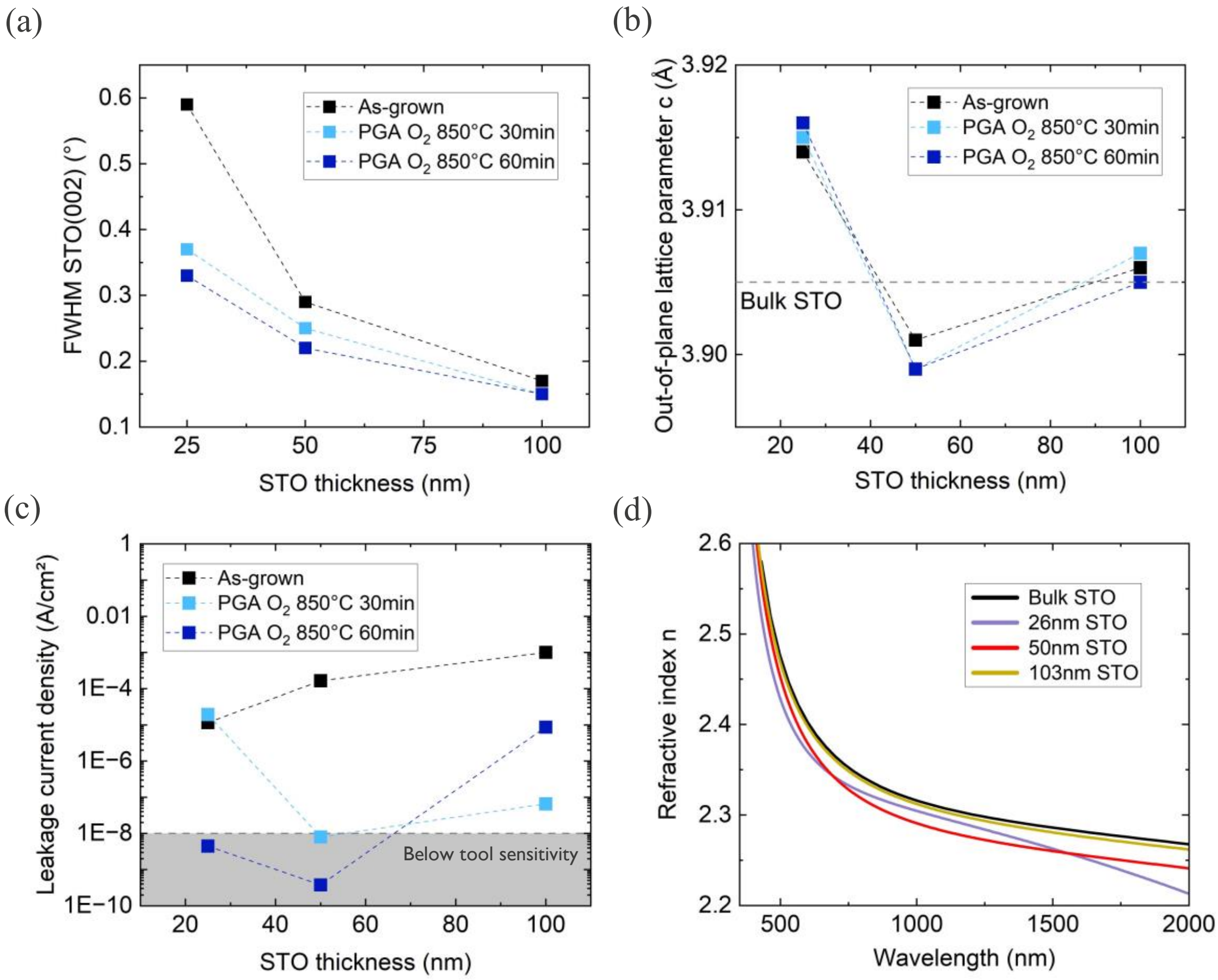

*Figure 8: Thickness dependence of stoichiometric STO films. (a) FWHM of the STO (002) rocking curve as a function of STO thickness before and after PGA. (b) Out-of-plane lattice parameter as a function of STO thickness before and after PGA. (c) Leakage current density at 1E+7 V/m as a function of STO thickness before and after PGA. (d) Spectrum of refractive index n for the three STO samples with different thickness after PGA compared to bulk STO* [38].

## IV Discussion

Growing high-quality crystalline STO on Si using MBE presents significant challenges, including the complexity of maintaining stoichiometry and ensuring abrupt interfaces. The results of our study on the impact of STO cationic stoichiometry on the layer crystallinity indicate that a perfect stoichiometry is required to grow layers with bulk-like lattice parameters.

For stoichiometric STO, in-situ annealing in $O_2$ was beneficial for structural and optical properties, as the as-grown layer already approximated the refractive index *n* of bulk STO.

Nevertheless, the electrical properties were still highly improved by additional ex-situ PGA in $O_2$ as the leakage current decreased several orders of magnitude because more oxygen vacancies were filled, which would otherwise facilitate charge carrier movement. However, the exceptionally low leakage current is not solely attributed to this; it also results from the increased thickness of the insulating interfacial $SiO_2$ layer, leading to an overall reduction in leakage current across the entire Cr/STO/$SiO_2$/Si/Pt stack.

Deviations in the Sr/Ti ratio will enlarge the STO unit cell due to more pronounced lattice distortions, as discussed in previous work [29,46–48]. AFM and XSEM results of the Sr-rich film show that PGA not only led to the formation of voids, cracks and a thick interfacial layer, but it also reduced the STO layer thickness from 75 nm to 62 nm. These observations can indicate that PGA probably caused redistribution of extra Sr atoms in the layer stack. Some excess of Sr could be baked off through the top interface [35]. Another part probably diffused towards the bottom interface where it could form an alloy with the interfacial $SiO_x$ layer, as excess Sr atoms can easily diffuse into $SiO_2$ at high temperatures [49]. The remaining layer left had therefore an improved stoichiometry as follows from the strongly reduced lattice parameter.

Next, the Ti-rich sample looks structurally similar to the stoichiometric one. The sample had a noticeable narrow (002) ω peak, which can indicate that incorporated excess $TiO_x$ affects the strain relaxation mechanism of STO [23,48]. Electrically and optically there is a clear degradation for the Ti-rich film and although after PGA the refractive index *n* recovers, the leakage current of the Ti-rich sample is still higher due to leakage pathways caused by excess Ti atoms and/or Sr vacancies [23,39,40].

The results of the thickness series clearly illustrate the relaxation behavior of STO layers on Si. For the thinnest STO film, the out-of-plane lattice parameter remained almost unchanged after PGA, indicating residual compressive strain. This difficulty in relaxing thin layers may stem from a higher density of defects due to the higher source oxidation during the first STO layers. This could influence the relaxation mechanism or the pronounced effects of simultaneous interface oxidation, which are more significant in thinner samples.

For the 50 nm STO layer, the out-of-plane lattice parameter was slightly below the bulk value, suggesting the presence of tensile strain. This strain likely arises during the cooling process due to the mismatch in thermal expansion coefficients between STO ($8.8 \times 10^{-6}$ $K^{-1}$) and Si ($2.6 \times 10^{-6}$ $K^{-1}$) [50]. In contrast, the as-grown and annealed 100 nm STO layers had an out-of-plane lattice parameter very close to the bulk value. Additional reciprocal space mapping (RSM) measurement around STO (103) diffraction confirmed that the 100 nm STO can be grown as fully relaxed material. These results indicate that further investigation is required to study and minimize induced strain in very thin STO films. This is particularly critical for applications where the dielectric permittivity of STO is a key parameter, as strain effects will influence the Curie temperature $T_C$ [4].

## V Conclusion

STO thin films with auspicious bulk-like properties require perfect stoichiometry. By employing RHEED as a real-time feedback mechanism, we effectively control cationic stoichiometry and mitigate the oxidation effects of the Sr source during STO epitaxy. Physical, electrical and optical characterization underscore the critical role of stoichiometry, as deviations from Sr/Ti = 1 lead to increased lattice parameters, leakage current and optical absorbance. Furthermore, high-temperature PGA treatments in $O_2$ can reduce the oxygen vacancy concentration, improving the STO insulating properties and optical transparency. These treatments also promote strain relaxation, improve crystallinity, and result in smoother surfaces for stoichiometric STO. High-quality STO films exceeding 100 nm are achieved with a rocking curve FWHM below 0.2°, closely matching the bulk refractive index. This study paves the way for using STO thin films as active materials in advanced devices for various applications, including energy storage and quantum information technology. Future work will focus on further increasing the STO thickness, measuring dielectric properties and optimizing the PGA conditions. Additionally, alternative measurement techniques will be explored to accurately determine optical losses, given the current sensitivity limitations of ellipsometry.

**Author contributions:** Conceptualization, A.B., M.B., J.V.d.V., C.M. and C.H.; methodology, A.B.; validation, A.B., C.M. and C.H.; investigation, A.B., A.U. and K.B.; resources, C.M. and C.H.; data curation, A.B.; writing—original draft preparation, A.B.; writing—review and editing, A.B., M.B., C.M. and C.H.; visualization, A.B.; supervision, C.M. and C.H.; project administration, C.M. and C.H.; funding acquisition, C.M. and C.H. All authors have read and agreed to the published version of the manuscript.

**Funding:** This research was funded by the Branco-Weiss Society and the European Research Council (ERC) under the European Union’s Horizon 2020 research and innovation program, grant number 864483 NOTICE and grant number 101042414 Q-AMP.

**Acknowledgments:** The authors extend their gratitude to the Material and Component Analysis department at IMEC for their invaluable assistance with sample characterization. Special recognition is given to Paola Favia for her guidance on interpreting the TEM results, Johan Meersschaut for executing the RBS investigation, Danielle Vanhaeren and Luca Mana for AFM results and Robert Gehlhaar for his support in constructing an ellipsometry model. A heartfelt thank you goes to Hans Costermans and Kevin Dubois for all assistance with the MBE cluster tool.

**Conflicts of Interest:** Authors Andries Boelen, Marina Baryshnikova, Anja Ulrich, Kamal Brahim, Christian Haffner and Clement Merckling are employed by the company IMEC. The authors declare that the research was conducted in the absence of any commercial or financial relationships that could be construed as a potential conflict of interest.

## Appendix A: Supporting figures

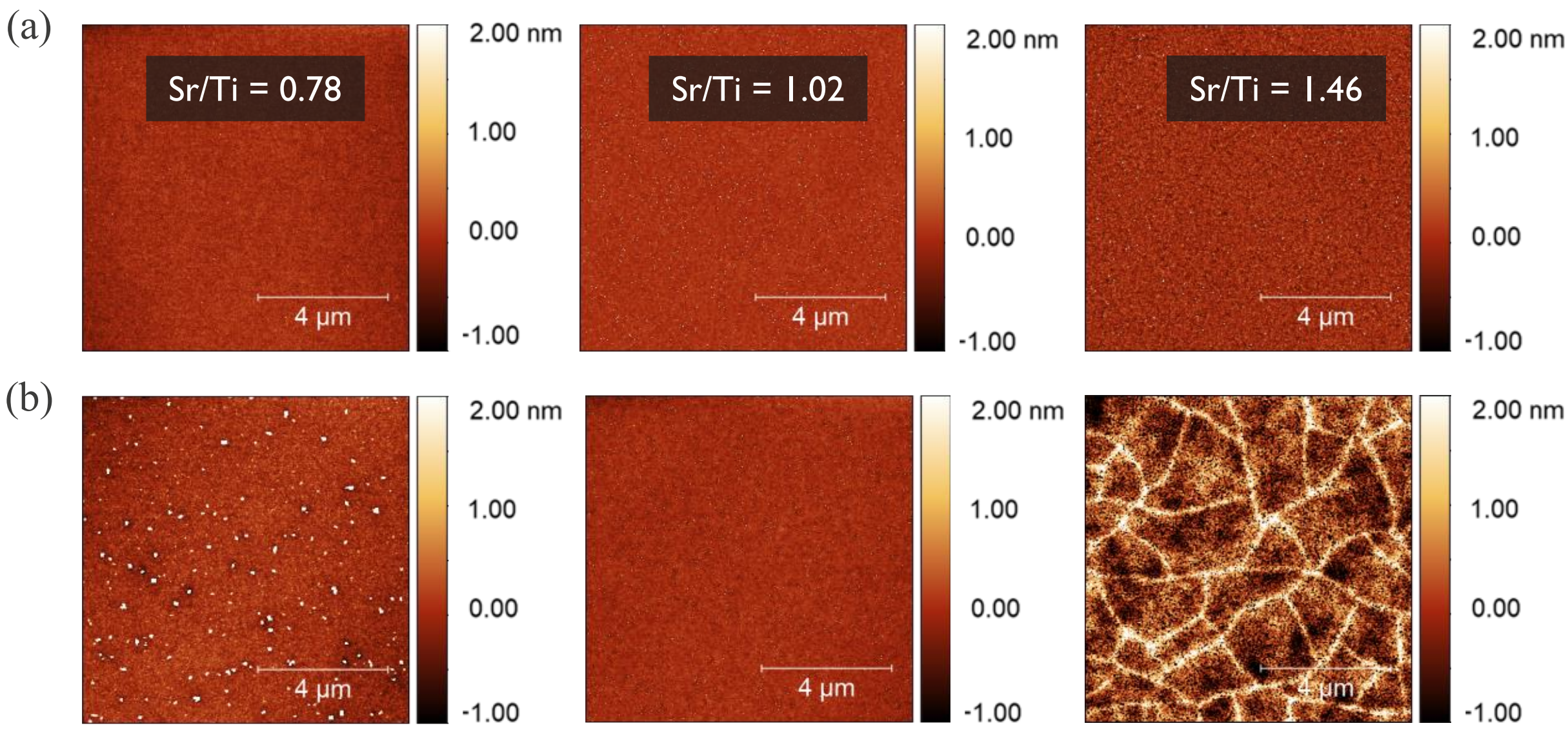

*Supplementary Figure 1: 10 x 10 µm AFM images for the three STO samples with varying cationic stoichiometry. (a) As-grown STO samples all show a smooth surface ($R_q$ < 0.4 nm). (b) STO after PGA 850°C 60 min. Large particle formation causes the surface roughness of Ti-rich STO to increase, whereas deep cracks and pinholes severely degrade the surface of Sr-rich STO.*

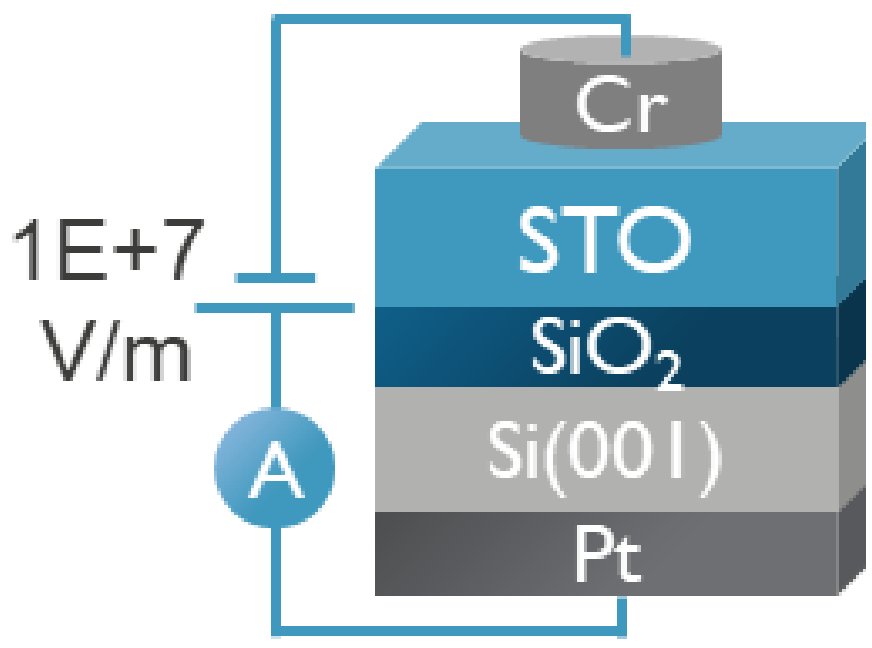

*Supplementary Figure 2: Schematic of the Cr/STO/$SiO_2$/Si/Pt stack used for I-V measurements applying 1E+7 V/m. The circular top contact has a radius of 200 µm.*

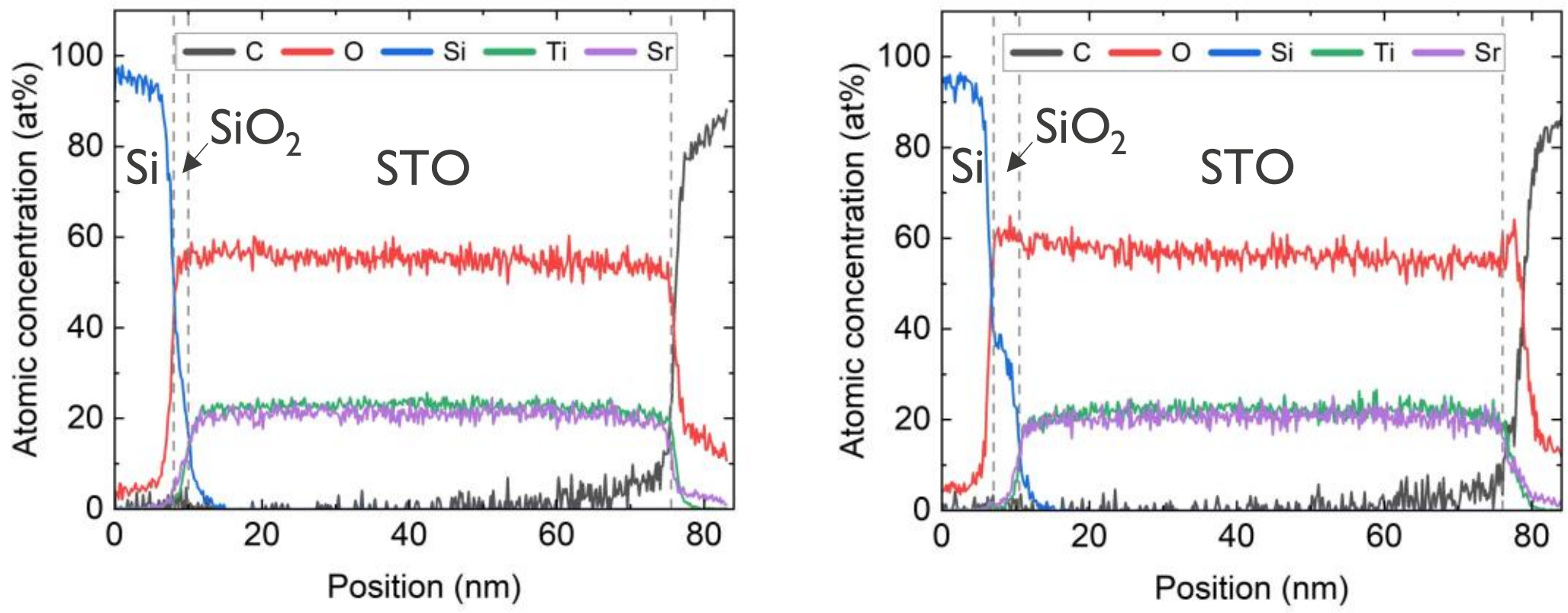

*Supplementary Figure 3: EDS corresponding to TEM in Figure 7 before (left) and after PGA (right). Both Sr and Ti atomic concentrations are constant throughout the film and the interfacial $SiO_2$ layer increases after PGA.*

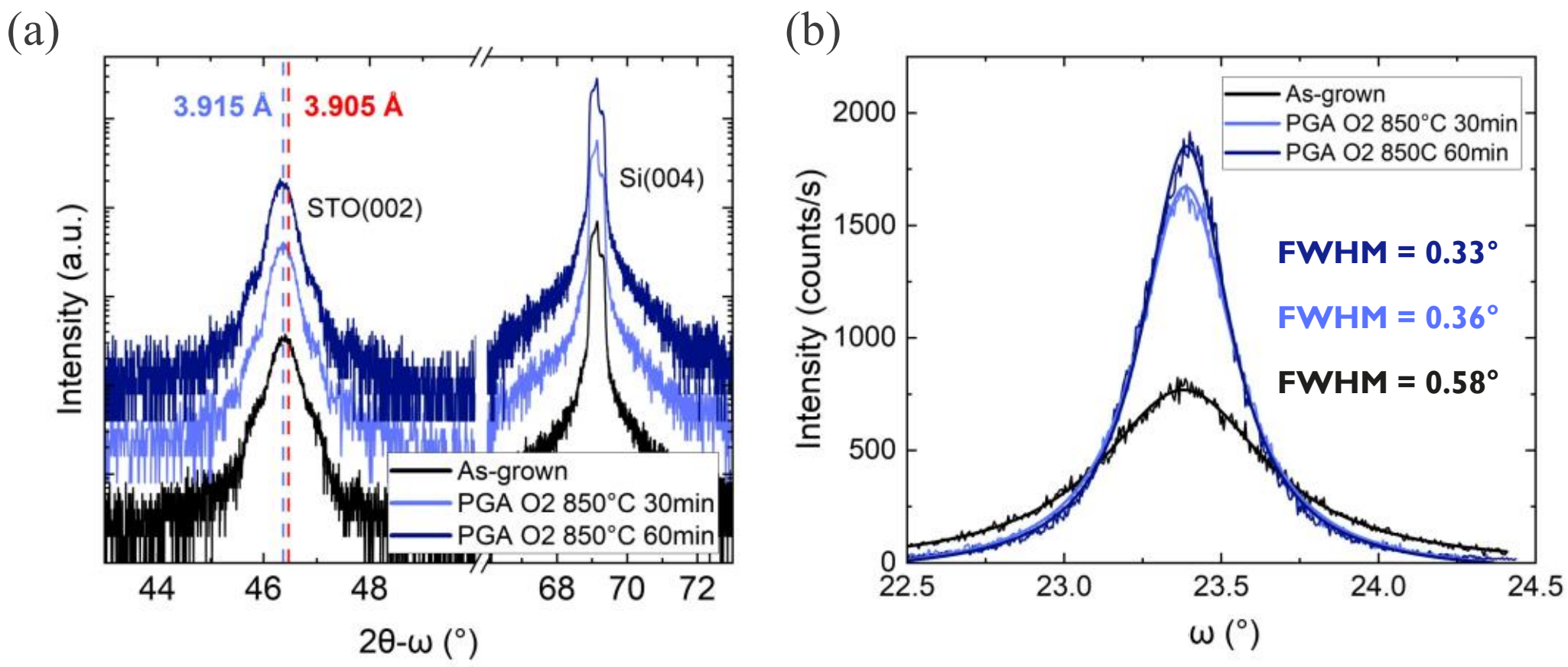

*Supplementary Figure 4: As an example of the XRD results, the 26 nm stoichiometric STO film is shown before and after PGA. (a) 2θ-ω scan showing the STO (002) and Si (004) diffraction peaks. The out-of-plane lattice parameter of the film after PGA for 30 min (light blue) and bulk (red) are indicated by dashed vertical lines. (b) ω scan shows the decreasing FWHM after PGA, indicating improved crystallinity.*

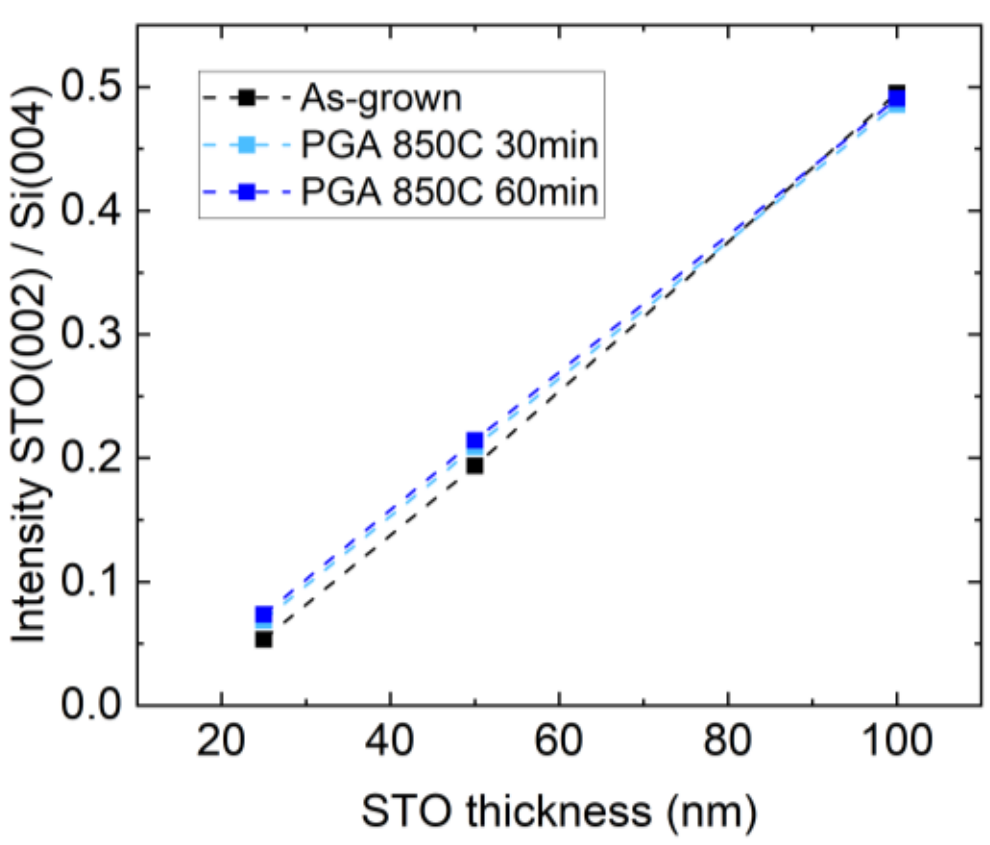

*Supplementary Figure 5: XRD intensity of the STO (002) diffraction peak as a function of STO thickness before and after PGA showing a linear increase as expected.*

**Appendix B: Ellipsometry fitting**

The model used for fitting ellipsometry data consists of five Tauc-Lorentz oscillators and includes both an intermediate silicon-oxide $SiO_x$ layer and a $SiO_2$ layer between the STO and Si to model the silicon oxidation. AFM surface roughness from the 10 x 10 µm scan was used to define the range for the fitted roughness, with a margin of ~ 0.1 nm from the AFM value.

The mean squared error (MSE) of the fitting is calculated by the CompleteEASE software using the following equation:

$$MSE = \sqrt{\frac{1}{3n - m}\sum_{i=1}^{n}\left[\left(\frac{N_{E_i} - N_{G_i}}{0.001}\right)^2 + \left(\frac{C_{E_i} - C_{G_i}}{0.001}\right)^2 + \left(\frac{S_{E_i} - S_{G_i}}{0.001}\right)^2\right]}$$

where $n$ is the number of wavelengths, $m$ is the number of fitting parameters, and $N = \cos(2\Psi)$, $C = \sin(2\Psi \mathrm{i}) \cos(\Delta)$, $S = \sin(2\Psi) \sin(\Delta)$. Subscript $E$ and $G$ denote measured and generated data, respectively.

For all our samples, MSE values are below 8, indicating a good fit. As-grown samples all have a fitted $SiO_2$ layer below 2.4 nm, whereas after PGA it is between 3.3 nm and 4.8 nm, in accordance with TEM. Only for the annealed Sr-rich STO the fit is worse due to the thick ~ 40 nm interfacial layer and high defectivity, resulting in an MSE of 25.4.